# Interpreting non-random signatures in biomedical signals with Lempel-Ziv complexity

*Radhakrishnan Nagarajan
Department of Biostatistics, University of Arkansas for Medical Sciences

Janusz Szczepanski
Institute of Fundamental Technological Research, Polish Academy of Sciences

Eligiusz Wajnryb
Institute of Fundamental Technological Research, Polish Academy of Sciences

Biostatistics, COPH/UAMS, Room 3234
4301 W Markham, Slot 781
Little Rock, AR 72205-7199
Tel: 501-526-6734
Fax: 501-526-6729
Email: nagarajanradhakrish@uams.edu

**Abstract**

Lempel-Ziv complexity (LZ) [1] and its variants have been used widely to identify non-random patterns in biomedical signals obtained across distinct physiological states. Non-random signatures of the complexity measure can occur under nonlinear deterministic as well as non-deterministic settings. Surrogate data testing have also been encouraged in the past in conjunction with complexity estimates to make a finer distinction between various classes of processes. In this brief letter, we make two important observations (1) Non-Gaussian noise at the dynamical level can elude existing surrogate algorithms namely: Phase-randomized surrogates (FT) amplitude-adjusted Fourier transform (AAFT) and iterated amplitude adjusted Fourier transform (IAAFT). Thus any inference nonlinear determinism as an explanation for the non-randomness is incomplete (2) Decrease in complexity can be observed even across two linear processes with identical auto-correlation functions. The results are illustrated with a second-order auto-regressive process with Gaussian and non-Gaussian innovations. AR (2) processes have been used widely to model several physiological phenomena, hence their choice. The results presented encourage cautious interpretation of non-random signatures in experimental signals using complexity measures.

# 1. Introduction

Several complexity measures including Lempel-Ziv complexity [1] and its variants [2] have been used widely to quantify the regularity and extent of randomness in data sampled from physical and biological systems [3-15]. Such systems are undoubtedly nonlinear feedback systems corrupted with noise at the *dynamical* $(\in_t)$ and *measurement* $(\eta_t)$ levels. While the former $(\in_t)$ is a feedback process coupled to the systems dynamics e.g. $x_t = f(x_{t-1}) + \in_t$, the latter $(\eta_t)$ is additive and acts externally e.g. $y_t = f(y_{t-1}); x_t = y_t + \eta_t$. Subsequently, these are mapped onto the observed value through a measurement device with an associated *transfer function*. The step-wise procedure mapping the true dynamics onto the observed value can be represented by a schematic diagram [16]. Complexity measures have been successfully used to detect possible non-random structure and discriminate different physiological states of activity [3-15]. In order to make a finer distinction of the observed non-random signatures, surrogate testing [17, 18] have been used in conjunction with complexity measures [13-16]. This includes complexity measures such as LZ complexity and its extensions ($\gamma$) [12-15].

In general, LZ complexity measures the rate of generation of new patterns along a sequence and in the case of ergodic processes is closely related to the entropy rate of the source [7]. A decrease in the complexity ($\gamma$) has been attributed to presence of non-random patterns in the given data. While there have been attempts to argue in favor of nonlinear deterministic signatures (possibly chaotic) patterns in physiological data [14 15], recent studies have pointed out that a decrease in complexity can occur in the case of nonlinear deterministic as well as non-deterministic processes [13]. Thus, any conclusion on the nature of the process based solely on the complexity measure is unhelpful.

Subsequently, surrogate testing was proposed in conjunction with complexity estimates for finer classification of the dynamics [9, 12, 13].

In this brief letter we make two important observations (1) we show that the existing surrogate algorithms used in conjunction with complexity measure may not be sufficient to draw conclusions on the existence of nonlinear deterministic signatures [9, 10]. More importantly, we show that the null hypotheses can be rejected across FT, AAFT and IAAFT surrogate algorithms even in the case of simple linear processes. Thus any argument in favor of nonlinear determinism based on the results of surrogate testing in conjunction with complexity measure is incomplete. (2) A decrease in complexity can be observed even across linear processes with identical auto-correlation functions. Thus a decrease in complexity need not necessarily imply a change of the auto-correlation function. The above results are demonstrated on second-order auto-regressive process, AR (2) with Gaussian and non-Gaussian innovations. The present study is in conjunction with our recent efforts to understand the impact of non-Gaussian innovations on surrogate testing [19] and the long-standing interest in understanding the pitfalls of surrogate algorithms [20-23]

## 2. Methods

*2.1 Second-order auto-regressive process AR(2)*

Auto-regressive modeling has been used successfully to capture the correlation and spectral properties of biomedical signals [20]. Recent studies, [21-23] have demonstrated the usefulness of second-order auto-regressive process AR (2) in modeling physiological tremor and the fact that it represents a well-defined elementary stochastic process are some of the reasons for its choice in the present context.

An AR (2) process is given by the expression

$$x_t = \alpha_1 x_{t-1} + \alpha_2 x_{t-2} + \epsilon_t \quad \ldots\ldots\ldots\ldots\ldots\ldots\ldots\ldots\ldots\ldots\ldots\ldots\ldots\ldots\ldots (1)$$

where $\alpha_1$ and $\alpha_2$ are the process parameters and $\epsilon_t$ is an independent and identically distributed (i.i.d) innovations. The parameters were fixed as $\alpha_1 = 0.8$ and $\alpha_2 = -0.5$ and correspond to a stationary AR (2) process. The choice of these parameters is encouraged by recent articles which used identical parameters to study physiological tremors [22, 23]. We consider two instances of dynamical noise, namely: $\epsilon_t^1 = \eta_t$ (normally distributed) and $\epsilon_t^2 = e^{\eta_t}$ (log-normally distributed) where $\eta_t$ is zero-mean, unit-variance i.i.d Gaussian noise. By generating $\epsilon_t^2$ as nonlinear transform Gaussian $\epsilon_t^1$ facilitates direct comparison of their corresponding AR (2) realizations (1). In subsequent discussion (Sec. 3), these shall be referred to as *paired observations*. We define abbreviations AWGN (additive white Gaussian noise) and AWNGN (additive white non-Gaussian noise) as follows.

**AWGN**: AR (2) process (1) generated with parameters $\alpha_1 = 0.8$ and $\alpha_2 = -0.5$ and Gaussian innovations $\epsilon_t^1$ as described above.

**AWNGN**: AR (2) process (1) generated with parameters $\alpha_1 = 0.8$ and $\alpha_2 = -0.5$ and non-Gaussian innovations $\epsilon_t^2$ as described above.

It should be noted that AWGN and AWNGN are *linearly correlated process* with identical auto-correlation function irrespective of the choice of the dynamical noise.

*2.2 Lempel-Ziv complexity*

Lempel and Ziv [1] proposed an algorithm to generate a given sequence using two fundamental operations, namely: *copy* and *insert* by parsing it from left to right. The Lempel-Ziv complexity $c(n)$ of a sequence of length $n$ is given by the shortest sequence generated using the copy and insert operation that can generate the given sequence. This shortest sequence is random although the sequence it generates need not necessarily be random. Thus any apparent patterns or correlations in a given sequence renders its complexity $c(n)$ lesser than that of a random sequence. More importantly, the asymptotic behavior of $c(n)$ in the case of uniformly distributed symbols is given by $b(n) = \dfrac{n}{\log n}$. Subsequently, $c(n)$ is normalized to $b(n)$ resulting in $\gamma = \dfrac{c(n)}{b(n)}$. The above definition of b(n) implicitly assumes the Shannon entropy of the sequence to be unity. In the present study, we consider a binary partition about the median which implicitly renders the Shannon entropy to be unity. Kaspar and Schuster [2] explored the choice of the LZ algorithm to quantify complex dynamical behavior. A detailed description of their algorithm along with its implementation for determining the complexity of a binary sequence can be found elsewhere [2]. A simple example is illustrated below for completeness a more formal definition can be found elsewhere. Prior to the discussion of the example we introduce the notation $v(s)$ in the following example corresponds to the vocabulary set [1, 2] or the set of *words* that can be generated from $s$. Consider $s = 00$, then $v(s)$ represents all possible words that can be reconstructed from s when scanning from left to right, i.e. $v(s) = \{0, 00\}$. If the incoming bit is 1, it cannot be generated from $v(s)$, hence an *insert* is required. However, if the incoming bit is a 0, it can be generated from $v(s)$, hence only a *copy* is required. Each time an insert operation occurs a dot is placed in the appropriate position [1, 2]. Complexity $c(n)$ of a period 3 sequence $s = 001001001....$ is shown below.

(a) The first digit 0 is unknown hence have to be inserted resulting in $c(n) = 1$ and $s^* = 0\bullet$.

(b) Consider the second digit 0. Now $s = 0$, $q = 0$; $sq = 00$; $sq\pi = 0$; $q \in v(sq\pi)$; therefore copying is sufficient resulting in no change in the complexity i.e. $c(n) = 1$ and $s^* = 0\bullet 0$.

(c) Consider the third digit 1. Now $s = 0$, $q = 01$; $sq = 001$; $sq\pi = 00$; $q \notin v(sq\pi)$; therefore insertion is required resulting in $c(n) = 2$ and $s^* = 0\bullet 01\bullet$.

(d) Consider the fourth digit 0: $s = 001$; $q = 0$; $sq = 0010$; $sq\pi = 001$; $q \in v(sq\pi)$; therefore copying is sufficient resulting in $c(n) = 2$ and $s^* = 0\bullet 01\bullet 0$.

(e) Consider the fifth digit 0: $s = 001$; $q = 00$; $sq = 00100$; $sq\pi = 0010$; $q \in v(sq\pi)$; therefore copying is sufficient resulting in $c(n) = 2$ and $s^* = 0\bullet 01\bullet 00$.

(f) Consider the fifth digit 1: $s = 001$; $q = 001$; $sq = 001001$; $sq\pi = 00100$; $q \in v(sq\pi)$; therefore copying is sufficient resulting in $c(n) = 2$ and $s^* = 0\bullet 01\bullet 001$.

Subsequent additions does not change $c(n)$. Since the sequence $s^*$ does not end in a dot (.) we add one to the resulting $c(n)$, resulting in $c(n) = 3$ for the given period three sequence s = 001001001…..

It is important to recall that the objective of the present study is to understand the relative change in the complexity between the empirical sample and the surrogate counterparts. AAFT and IAAFT surrogates retain the distribution of the empirical sample in the surrogate realizations by their very construction; hence the Shannon entropy is preserved. In the case of the FT surrogates significant discrepancy in the distribution can be observed when the normality assumption of the empirical sample is violated. However, partitioning about the median implicitly renders the Shannon entropy to be unity. Thus the normalizing factor $b(n)$ does not really play an important role in the present context.

*2.3 Surrogate testing*

Surrogate testing is useful in determining the nature of the process generating the given *empirical sample*. The term empirical sample reflects the fact that the given single realization is sufficient to capture the process dynamics. Such an assumption is especially valid for ergodic processes [24]. The essential ingredients of surrogate testing include: (a) *null hypothesis* (*Ho*), (b) *discriminant statistic* (*D*) (c) *surrogate algorithm* (*A*). The surrogate algorithm is designed so as to retain certain *essential* statistical properties of the empirical sample as dictated by the null hypothesis. The discriminant statistic is chosen so that its estimate between the given empirical sample and the surrogate realization shows considerable discrepancy when the null hypothesis is violated. An incomplete list of major contribution to this exciting area of research includes [12, 17, 18]. Three widely used surrogate algorithms include (i) *Phase-randomized surrogates* (FT), (ii) *amplitude adjusted Fourier transform* (AAFT) and (iii) *iterated amplitude adjusted Fourier transform* (IAAFT). The most elementary however is random shuffled surrogates which address the null that the given data is generated by an i.i.d process. Since the processes to be considered are correlated processes, we expect the null to be rejected in the case of random shuffled surrogates. Therefore, we do not discuss the results of random shuffled surrogates. FT surrogates address the null that the given empirical sample is generated by a linearly correlated process. Thus the power-spectrum of the empirical sample is retained in the surrogate realizations. It should be noted that the power-spectrum of a stationary linear process is related to its auto-correlation function by Wiener-Khinchin theorem and the parameters of a linear process can be estimated from their auto-correlation function (Yule-Walker equations) [24]. Hence, retaining the auto-correlation function completely specifies the process. Often, the underlying dynamics is mapped onto an observed value through a transducer or measurement device with a nonlinear transfer function. Such nonlinear transforms deemed trivial and assumed to be static, invertible transforms. These in turn renders the distribution of the signal to be non-Gaussian. AAFT surrogates address the null that the given data is generated by a static, invertible nonlinear transform of a linearly correlated noise. The objective is to retain the amplitude distribution as well as the power-spectrum in the surrogate realization. IAAFT surrogate is a significant improvement over AAFT surrogates and retains the amplitude distribution and the power-spectrum to a greater accuracy.

Since the analytical form of the AR (2) process (1) is known, we directly compare the complexity ($\gamma$) obtained on several independent empirical realizations to those obtained on their FT, AAFT, IAAFT surrogate counterparts. Such a comparison is accomplished using parametric (ttest) and non-parametric (wilcoxon-ranksum) tests [25] at significance level ($\alpha = 0.05$). While the parametric determines statistical significance based on the true values, non-parametric test uses the ranks as opposed to the true values. Unlike parametric test, non-parametric test does not assume normal distribution of the complexity ($\gamma$) estimates. The null hypothesis addressed is that there is no significant difference in the complexity estimates obtained on the empirical samples and their surrogate counterparts. Rejecting the null hypothesis implies significant difference in the complexity estimates and accompanied by a low p-value ($< 0.05$).

## 3. Results

In order to establish the fact that complexity ($\gamma$) can differ considerably across two distinct processes with the same auto-correlation function, we compared its estimate on 100 independent AWGN and AWNGN realizations by partitioning about the median, Fig. 1. While AWGN and AWNGN have identical auto-correlation functions, complexity ($\gamma$) of AWNGN was significantly lesser than those estimated on AWGN. This illustrates the fact that the complexity measure is sensitive ($\gamma$) to the choice of innovations (Gaussian or non-Gaussian) across linear processes with identical auto-correlation function. It is important to note that the LZ complexity is a nonlinear measure that defies the principle of superposition. Nonlinear measures in general are sensitive to higher order correlations in data which can arise due to nonlinearity or non-Gaussianity.

Empirical samples from AWGN and AWNGN processes were generated (N = $2^{12}$ samples) after discarding the initial transients. For the AWGN process, FT, AAFT and IAAFT surrogates retain the amplitude distribution (Figs. 2a-2c) as well as the power-spectrum (Figs. 3a-3c) of the empirical sample. The conformity of the distribution between the empirical sample and their corresponding surrogate counterpart is revealed a straight line along the diagonal of the quantile-quantile (QQ) plots, Figs. 2. However, the above is not true in the case of AWNGN process. While the power-spectrum of AWNGN is retained in its FT surrogate, Fig. 3d, the distribution is not, Fig. 2d. The discrepancy in the distribution is reflected by the marked deviation from the diagonal line, Fig. 2d. Although AWNGN is linear process, FT surrogates are not faithful in preserving the amplitude distribution, hence cannot be used for statistical inference of AWNGN processes. The power-spectrum and the amplitude distribution of AWNGN and its AAFT surrogate are shown in Figs. 3e and 2e respectively. While the distribution of AWNGN is preserved in its AAFT surrogate, Fig. 2e, there is notable discrepancy in the power-spectrum, Fig. 3e. Therefore, AAFT surrogates might not be appropriate for reliable statistical inference of AWNGN process. The distribution and the power-spectrum of AWNGN and its IAAFT surrogate are shown in Figs. 2f and 3f respectively. Unlike FT and AAFT surrogates, the power-spectrum as well as the distribution of AWNGN is faithfully retained only in the case of IAAFT surrogates.

The distribution of the complexity ($\gamma$) obtained on 100 independent realizations of the AWGN and their corresponding FT surrogate realizations obtained about partition is shown in Fig. 4a. For AWGN, there is a significant overlap between the distributions of ($\gamma$) obtained on the empirical sample and its FT surrogate, Fig. 4a. As expected, both parametric (ttest) and non-parametric (wilcoxon-ranksum) correctly failed to reject the null that there is significant difference in the complexity estimate on AWGN and its FT surrogate counterpart at ($\alpha = 0.05$). A similar analysis of AWNGN process and its FT surrogates about median partition is shown in Fig. 4d. Unlike AWGN, distribution of the complexity estimates on AWNGN and those of its FT surrogates were well separated. Parametric and non-parametric tests spuriously rejected the null at ($\alpha = 0.05$). A similar analysis of AWGN and AWNGN processes and their AAFT surrogates about the median partition is shown in Figs. 4b and 4e, respectively. As expected, parametric and non-parametric test correctly failed to reject the null in the case of AWGN at ($\alpha = 0.05$), Fig. 4b. However, the null was spuriously rejected in the case of AWNGN process, Figs. 4e. Similar results were obtained with IAAFT surrogates. Parametric and non-parametric tests correctly failed to reject the null in the case of AWGN Fig. 4c at ($\alpha = 0.05$). However, the null was spuriously rejected in the case of AWNGN Fig. 4f.

From the above discussion, it is important to note that the complexity ($\gamma$) is sensitive to the distribution of the noise term. It is equally important to note that all three surrogate algorithms rejected the null of linear process in the presence of non-Gaussian innovations even for a second-order auto-regressive process. This was demonstrated across partitioning with respect to median, Figs. 4d-4f. This was shown even across IAAFT surrogates. More importantly, rejecting the null hypothesis using complexity ($\gamma$) does not necessarily imply presence of even static nonlinearity in a given process.

**4. Discussion**

The present study clearly demonstrates that complexity measure ($\gamma$) in conjunction with FT, AAFT and IAAFT surrogate algorithms might not be useful in determining the nature of non-randomness in a physiological signal. Thus any conclusion on nonlinear deterministic patterns as a possible explanation to the observed non-random signatures is incomplete. Linear processes are usually specified by their auto-correlation functions. In the present study, we showed that a difference in complexity can occur even across linear processes with identical auto-correlation functions.

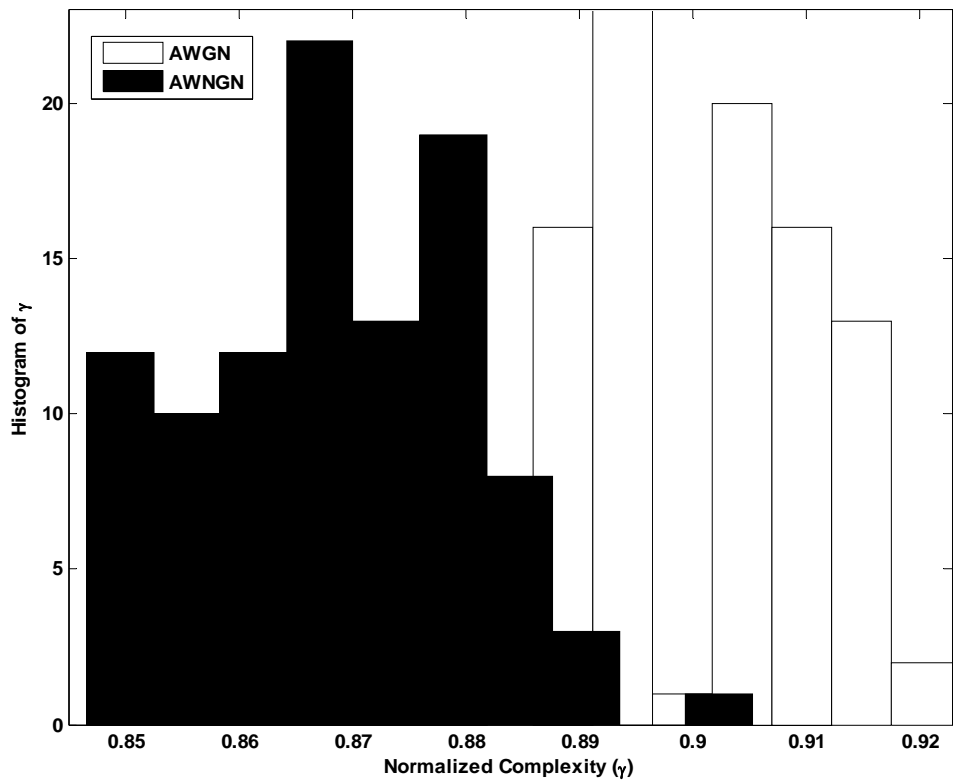

**Figure 1** Histogram of the normalized complexity ($\gamma$) estimates for 100 independent AR (2) processes with Gaussian (white bars) and non-Gaussian (black bars) innovations generated by partitioning about the median (N = $2^{15}$ samples).

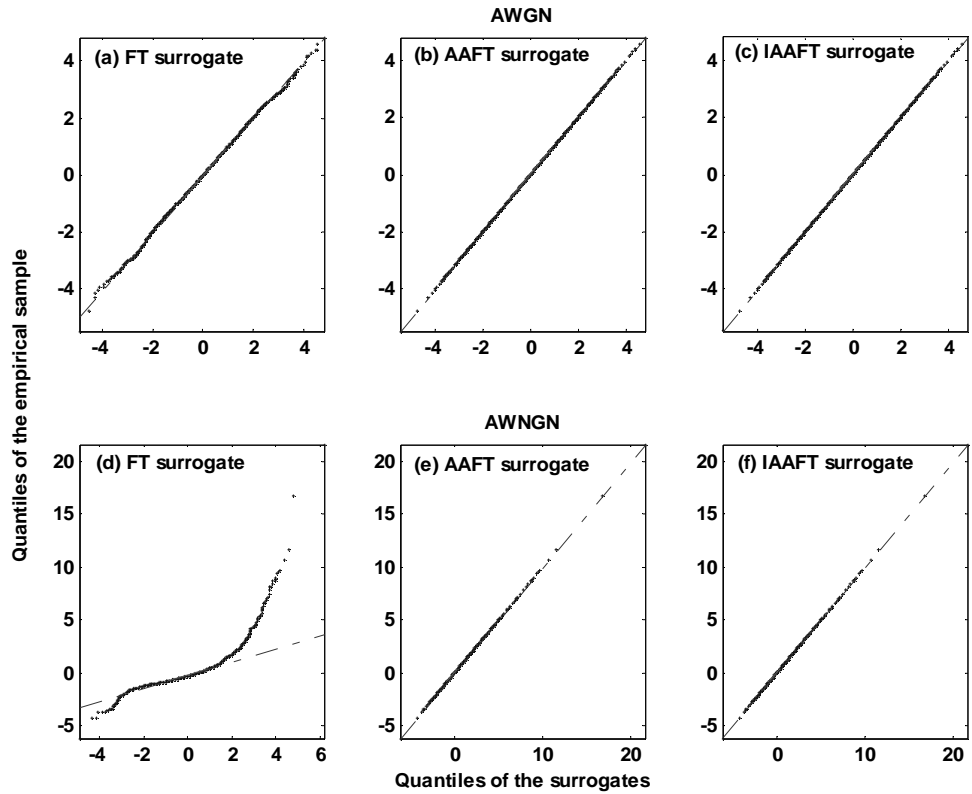

**Figure 2** QQ plot of the empirical samples against their FT, AAFT and IAAFT surrogate counterparts for the AWGN (a-c) and AWNGN (d-f) processes (N = 4096 samples). The diagonal dashed line represents the case where the distributions are identical.

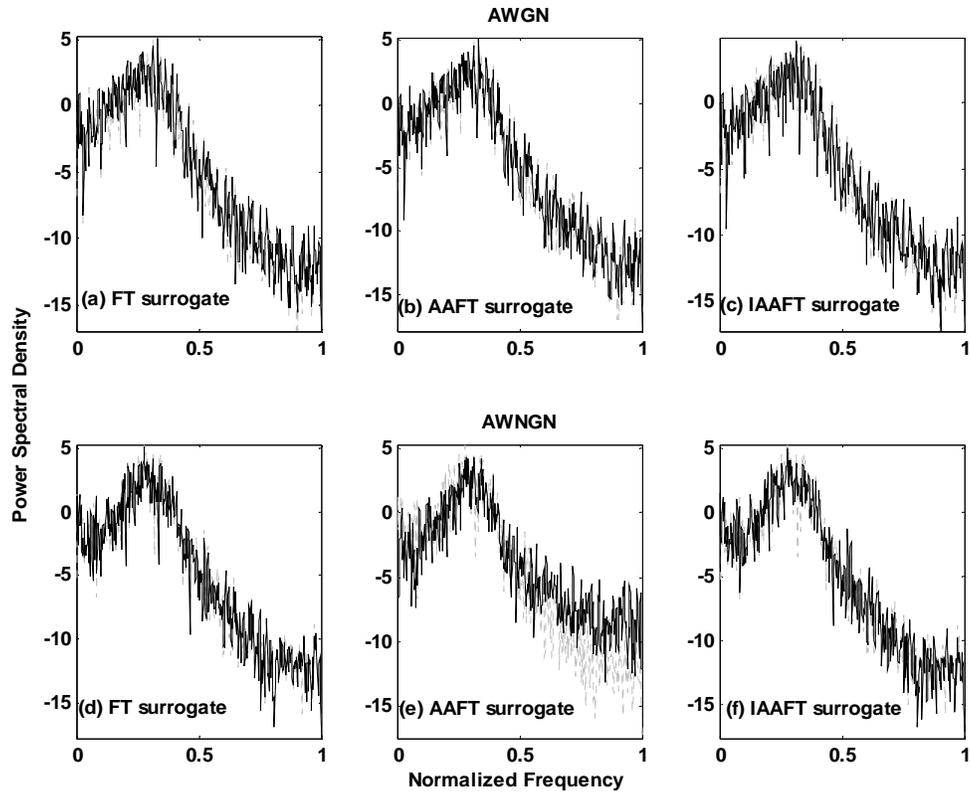

**Figure 3** Welch power-spectral density estimates of the empirical samples and their corresponding FT, AAFT and IAAFT surrogate counterparts for AWGN (a-c) and AWNGN (d-f) processes (N = 4096 samples).

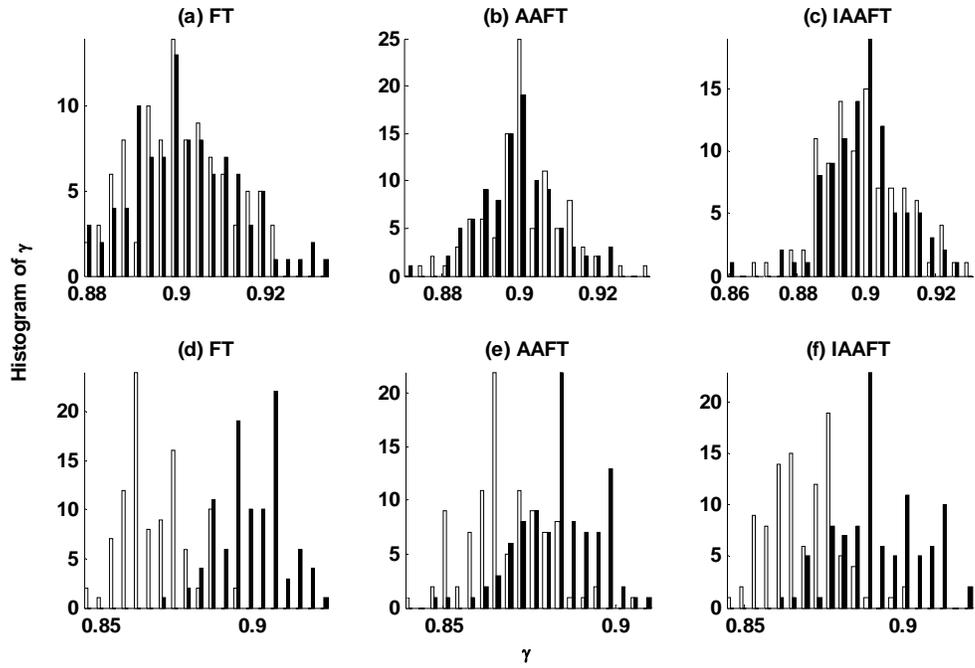

**Figure 4** Histogram of the normalized complexity ($\gamma$) obtained on the empirical samples (white bars) and the FT, AAFT and IAAFT surrogates (black bars) for the AR(2) process (2) with Gaussian (a, b and c) and non-Gaussian innovations (N = 4096 samples) generated by partitioning about the median.